\documentclass[fleqn,usenatbib]{mnras}

\usepackage{graphics,epsf}
\usepackage{amsmath}                
\usepackage{amsfonts}               
\usepackage{amssymb}                
\usepackage{epsfig}                 
\usepackage{graphicx}

\usepackage{xcolor}
\definecolor{redak}{rgb}{0.9,0.15,0.05}

\def \kms{~\rm{km~s^{-1}}}

\def \msyr{~\rm{M_{\odot}}~\rm{yr^{-1}}}

\def \K{~\rm{K}}

\def \AU{~\rm{au}}

\def \rmModot{~\rm{M_{\sun}}}
\def \rmRodot{~\rm{R_{\sun}}}
\def \rmLodot{~\rm{L_{\sun}}}

\title[Counteracting tidal circularization with the GEE]{Counteracting tidal circularization with the grazing envelope evolution}

\author[A. Kashi and N. Soker]{
Amit Kashi$^{1}$\thanks{E-mail: \href{mailto:kashi@ariel.ac.il}{kashi@ariel.ac.il}}
and
Noam Soker$^{2,3}$\thanks{E-mail: \href{mailto:soker@physics.technion.ac.il}{soker@physics.technion.ac.il}}
\\
$^{1}$Department of Physics, Ariel University, Ariel, POB 3, 40700, Israel \\
$^{2}$Deparment of Physics, Technion, Haifa 3200003, Israel\\
$^{3}$Guangdong Technion Israel Institute of Technology, Shantou, Guangdong Province, China
\\
}

\date{Accepted 2018 August 1. Received 2018 August 1; in original form 2018 July 11}

\pubyear{2018}

\begin{document}
\label{firstpage}
\pagerange{\pageref{firstpage}--\pageref{lastpage}}
\maketitle

\begin{abstract}
We show that substantially enhanced mass loss at periastron passages, as is expected in the grazing envelope evolution (GEE), can compensate for the circularization effect of the tidal interaction in binary systems composed of an asymptotic giant branch (AGB) star and a main sequence secondary star.
By numerically integrating the equations of motion we show that under our assumptions the binary system can maintain its high eccentricity as the AGB star evolves toward the post-AGB phase. Our results can explain the high eccentricity of some post-AGB intermediate binaries (post-AGBIBs), i.e., those with an orbital periods in the range of several months to few years.
In the framework of the GEE, the extra energy to sustain a high mass loss rate comes from the accretion of mass from the giant envelope or its slow wind onto a more compact secondary star. The secondary star energizes the outflow from the AGB outer envelope by launching jets from the accretion disk.
\end{abstract}

\begin{keywords}
stars: AGB and post-AGB ---
(stars:) binaries: close ---
stars: jets ---
stars: kinematics and dynamics ---
stars: mass-loss ---
accretion, accretion disks
\end{keywords}

\section{INTRODUCTION}
\label{sec:intro}

There is a group of post-asymptotic giant branch (AGB) binary systems with properties that are in dissension with traditional calculations of binary stellar evolution. They have typical orbital periods in the range of about one hundred days to several years, and the eccentricity is relatively large, from circular orbits and up to $e \simeq 0.6$
(e.g., \citealt{Gorlovaetal2014, VanWinckeletal2014, Manicketal2017, Kluskaetal2018}).

The first puzzle is their orbital separation. Binary stellar evolutionary studies that include only mass loss, mass transfer, tidal interaction, and common envelope evolution (CEE), usually lead to either a larger final orbital separation due to mass loss, $a_{\rm f} \gg 1 \AU$, or to a much smaller orbital separation as a result of CEE, $a_{\rm f} \ll 1 \AU$. According to these traditional studies there should be a gap around $a_{\rm f} \approx 1 \AU$ in the distribution of orbital separations of post-AGB binaries (e.g., \citealt{Nieetal2012}).  
 
The existence of the above post-AGB intermediate binaries (post-AGBIBs) in the traditional orbital separation gap raises the possibility that another process plays a significant role in the evolution of post-AGBIBs.  
One such process to explain post-AGBIBs is the grazing envelope evolution (GEE; \citealt{Soker2017SNIIb}). According to the GEE, the more compact secondary star grazes the envelope of the giant star and accretes mass from its outer envelope or from its wind acceleration zone. Because of the orbital motion the gas is accreted with large enough specific angular momentum to form an accretion disk around the secondary star, and the accretion disk launches jets. 
The system enters a GEE phase when the jets can efficiently remove the outer layers of the envelope of the primary giant star and by that prevent or delay the onset of a common envelope phase \citep{SabachSoker2015, Soker2015, Soker2016a, Shiberetal2017, ShiberSoker2018, AbuBackeretal2018, LopezCamaraetal2018}. It is this rapid mass removal that leaves the orbital separation of post-AGBIBs to be about the maximum radius the primary star has attained on the AGB. 

The envelope mass removal process operates in a negative feedback cycle \citep{Soker2016Rev}. If the jets remove too much gas from their vicinity the accretion rate decreases and so does the power of the jets (e.g., simulations conducted by \citealt{MorenoMendezetal2017}). The removal of gas acts to increase the orbital separation, and this in turn can reduce accretion rate onto the secondary star, unless tidal forces overtake and cause the formation of a common envelope phase. Further evolution of tidal forces and giant expansion can resume the cycle. 

The GEE scenario which helps explain these post-AGBIBs is also supported by observations that indicate that the secondary star in several post-AGBIBs launches jets (e.g., \citealt{Wittetal2009, Thomasetal2013, Gorlovaetal2012, Gorlovaetal2015, Bollenetal2017}), and it is quite likely that most post-AGBIBs launch, or have been launching jets \citep{VanWinckel2017}. Some post-red giant branch (RGB) intermediate binaries (post-RGBIBs) that have similar properties to post-AGBIBs (e.g., \citealt{Kamathetal2016}) might also experience the GEE. 

The secondary star need not be at the surface of the giant star, as it can also accrete mass from the acceleration zone of the wind.  
Both Roche lobe overflow (RLOF) and an accretion from an ambient gas (Bondi Hoyle Lyttleton type of accretion) contribute to the mass transfer process.
\cite{Harpazetal1997} discuss how when a close secondary star approaches periastron the extended envelope of the AGB star overflows its Roche lobe and flows toward the secondary star. The extended envelope that \cite{Harpazetal1997} refer to is the zone of the slowly moving outflow of matter (wind) before it reaches the escape velocity from the AGB star. \cite{PodsiadlowskiMohamed2007} later termed this process a wind-RLOF and simulated it \citep{MohamedPodsiadlowski2007}.
According to the GEE, when the density in the acceleration zone of the wind is sufficiently high such that the accretion rate is high enough, jets are launched by the secondary star and remove mass from the acceleration zone of the wind. 

The second puzzle is the high eccentricity of post-AGBIBs (e.g., \citealt{Manicketal2017}). Simple estimates indicate that at the orbital separation of post-AGBIBs tidal forces should have circularized the orbit during the AGB phase of the primary star. 
Several mechanisms can act against the tidal forces and increase the eccentricity of post-AGBIBs.
One such mechanism is a kick to the white dwarf (WD) as it is born \citep{Izzardetal2010}. The question here is what mechanism will give the required spin to the WD.
Another mechanism is the interaction of the binary system with a circumbinary disk (e.g., \citealt{Waelkensetal1996, Dermineetal2013}). Indeed Observations show circumbinary disk in many post-AGBIBs (e.g., \citealt{Watersetal1993, Bujarrabaletal2005, Bujarrabaletal2007, Bujarrabaletal2017}), and it is likely that all post-AGBIBs have or had such a circumbinary disk (e.g., \citealt{Bujarrabaletal2013, Bujarrabaletal2018}). The calculations of \cite{Dermineetal2013} are of resonant interaction of the binary system with its circumbinary disk that takes place only in the post-AGB phase. They do not consider the evolution during the AGB phase. Hence, they do not refer to the puzzle of the orbital separation, but do manage to account for most observed eccentricities of post-AGBIBs. 
The eccentricities they obtain are higher than what \cite{Soker2000e} estimated for the resonant interaction because they used a more favorable parameters for the circumbinary disk, i.e., higher disk mass and smaller inner disk radius. \cite{Rafikov2016} found that the mechanism of circumbinary interaction is unlikely to work as it requires a very massive disk and a too long disk life time. 
      
The mechanism to enhance eccentricity that is relevant to our study involves mass loss and mass transfer, in particular mass loss near periastron passages (e.g., \citealt{VanWinckeletal1995, Soker2000e, BonacicMarinovicetal2008, Davisetal2015}). \citealt{DosopoulouKalogera2016a} and \cite{DosopoulouKalogera2016b} conduct a general study of the role of mass transfer and mass loss on the eccentricity. They were not aiming specifically on post-AGBIBs. 
\cite{Vosetal2015} include in one model tidally-enhanced wind mass-loss, and in another model they include both phase-dependent RLOF and eccentricity pumping by the interaction of the binary system with the circumbinary disk. The results of \cite{Vosetal2015} do not reproduce the observed trend of higher eccentricities at higher orbital periods. The above papers, as well as others
(e.g., \citealt{Siessetal2014}), attributed the enhanced mass loss rate during periastron passages to gravitational effects of orbital motion. We here attribute the extra mass loss rate mainly to jets that are launched by the secondary star as it grazes the surface or the wind-acceleration zone of the primary giant star. The advantage of the GEE is that the accretion of mass on to the secondary star releases more energy to remove the envelope. 
 
\cite{Nieetal2017} come up with a completely different suggestion. They study binary systems where the primary is a red giant branch star (RGB) rather than a post-red giant star.
These systems also observe higher eccentricity than predicted by tidal circularization.
The solution \cite{Nieetal2017} propose in order to keep the orbits eccentric, is to reduce the rate of tidal circularization by a factor of $\approx 100$.
The reason \cite{Nieetal2017} do not consider enhanced mass loss rate is that their claim that such mechanisms lead to a substantial amount of circumbinary dust that is not observed for their red giant binaries. We do not have this constraint as we consider post-AGBIBs and post-RGBIBs, and the giant primary stars in these systems have already lost most of their envelope mass. Indeed, circumbinary material is observed around most of these systems.

In the present study we assume that during the AGB (or RGB) phase of the primary star the secondary star grazes its surface or the base of the acceleration zone of its wind. We assume that jets launched by the accretion disk around the secondary star remove mass from the primary star and/or from the acceleration zone of its wind at a high rate. We consider the usual theoretical value of the tidal interaction and do not reduce it. We focus on the eccentricity of the system and do not consider other processes, such as the formation of a circumbinary disk (for recent detail study of circular orbits of AGB binaries see, e.g., \citealt{Chenetal2017, Chenetal2018}). 

In section \ref{sec:evolution} we present the processes that modify the orbital parameters as the AGB evolves. In sections \ref{sec:Tidal-dominated_evolution} and \ref{sec:Periastron-enhanced} we present the results of an evolutionary model which takes into account these processes.
We summarize in section \ref{sec:summary}.

\section{The Orbital Evolution}
\label{sec:evolution}

We shall define $M_1$ and $M_2$ as the masses of the giant and the secondary star, respectively,
and the mass of the system as ${M=M_1+M_2}$.
We also define the reduced mass ${\mu = M_1M_2/(M_1+M_2)}$.
We will adequately follow \cite{Eglleton2006} in taking Newton's equation and introducing perturbing forces that modify the orbit and the orbital parameters.
Doing so, the orbital separation $\mathbf{r}$ varies according to 
\begin{equation}
\ddot{\mathbf{r}}(t) = -\frac{GM(t)\mathbf{r}(t)}{r^3(t)} + f_{\rm{QD}} + f_{\rm{TF}} + f_{\rm{M}}, 
\label{eq:rt}
\end{equation}
where $G$ is the gravitational constant and the other terms have the following meanings.
 
The perturbing force (acceleration) due to tidal friction, under the assumption that the perturbing star is a point mass, is
\begin{equation}
f_{\rm{TF}}=-\frac{9\sigma M_2^2 A^2}{2\mu r^{10}} \left[3 \mathbf{r}(\mathbf{r} \cdot \dot{\mathbf{r}}) + (\mathbf{r} \times \dot{\mathbf{r}} -\mathbf{\Omega} r^2) \times \mathbf{r} \right].
\label{eq:ftf}
\end{equation}
In the above equation $\mathbf{\Omega}$ is the angular velocity of the giant, and
\begin{equation}
A=2k_{\rm AM}R_1^5
\label{eq:A}
\end{equation}
is a constant related to the structure of the giant, where $k_{\rm AM}$ is the apsidal motion constant.
We will express the apsidal motion differently, using the internal structure constant $Q$ which satisfies
\begin{equation}
k_{\rm AM}=Q/[2(1-Q)].
\label{eq:k_AM}
\end{equation}
We also define a dissipation coefficient which is also related to the structure of the star
\begin{equation}
\sigma=\frac{2}{M_1 R_1^2 Q^2 t_{\rm visc}},
\label{eq:sigma}
\end{equation}
where $t_{\rm visc}$ is the viscosity time.
We also consider here the effect on the orbit of the quadrupole distortion (or equilibrium
tide) of the giant due to its rotation and to the presence of a secondary star
\begin{equation}
f_{\rm{QD}}= - \frac{A M_2}{\mu} \left[\frac{\Omega^2 \mathbf{r}}{2r^5} + \frac{3GM_2\mathbf{r}}{r^8} \right].
\label{eq:fqd}
\end{equation}
We note that in the above equation we omitted terms with $\mathbf{\Omega} \cdot \mathbf{r}$ as we assume that the rotation axis of the giant is perpendicular to the orbital plane.
The perturbing force (acceleration) due to mass loss and mass transfer from the giant to the secondary star is 
\begin{equation}
f_{\rm{M}}=
\dot m_t \left(\frac{1}{M_1(t)}-\frac{1}{M_2(t)} \right) \dot{\mathbf{r}}(t).
\label{eq:fm}
\end{equation}
The rates of change of the stellar masses are
\begin{equation}
\dot M_1=-\dot m_{l1}-\dot m_t ~;~~
\dot M_2=-\dot m_{l2}+\dot m_t ~;~~
\dot M = \dot M_1 + \dot M_2,
\label{eq:Mdot}
\end{equation}
where $\dot m_{l1}$ and $\dot m_{l2}$ are the mass loss rates to infinity
from the primary and the secondary, respectively,
and $\dot m_t$ is the rate of mass transferred from the primary to the secondary.
We shall use the integral of the mass loss rates over time
\begin{equation}
M_{\rm{ej}}= \int_0^t \dot m_{l1}\,dt  ~;~~
M_{\rm{acc}} =\int_0^t \dot m_t \,dt.
\label{eq:Mejacc}
\end{equation}

We then perform integration over time of equation \ref{eq:rt} to calculate $\mathbf{r}(t)$ .
The equation cannot be solved analytically and is therefore solved numerically.
The eccentricity $e(t) \equiv |\mathbf{e}(t)|$ is calculated according to
\begin{equation}
GM(t)\mathbf{e}(t) =
\dot{\mathbf{r}}\times(\mathbf{r}\times\dot{\mathbf{r}}) -
\frac{GM\mathbf{r}}{r}.
\label{eq:e}
\end{equation}
The Keplerian energy per unit reduced mass $\varepsilon(t)$ is calculated according to
\begin{equation}
\varepsilon(t) = \frac{1}{2} \dot{r}^2
- \frac{GM}{r},
\label{eq:energy}
\end{equation}
and then it is possible to calculate the semi-major axis
\begin{equation}
a(t) = - \frac{GM(t)}{2\varepsilon(t)},
\label{eq:a}
\end{equation}
and the orbital period
\begin{equation}
P(t)=2\pi\sqrt{\frac{a^3(t)}{GM(t)}}.
\label{eq:P}
\end{equation}

The parameter which is hard to determine is the viscosity time.
It is sensitive to the structure of the giant.
In order to calibrate this parameter for HD~44179 we take the
circularization time-scale, given by \cite{VerbuntPhinney1995} 
\begin{equation}
\begin{split}
\frac{1}{\tau_c} \equiv -\frac{d\ln{e}}{dt}
&= 1.7 f \left( \frac{T_{\rm eff}}{4500 \K} \right)^{4/3}
\left( \frac{M_{\rm env}}{\rmModot} \right)^{2/3}   \\
&\times \frac{\rmModot}{M} \frac{M_2}{M}
\frac{M+M_2}{M}
\left( \frac{R_1}{a} \right)^{8} \rm{yr^{-1}}      \\
\end{split}
\label{eq:tau_c}
\end{equation}
where $M_{\rm env}$ is the mass of the giant's envelope and $T_{\rm eff}$ is its effective temperature.

\section{Tidal-dominated evolution}
\label{sec:Tidal-dominated_evolution}

Motivated by the post-AGB binary system of the red rectangle, HD~44179 (\citealt{Cohenetal1975, Cohenetal2004, VanWinckel2004, Thomasetal2013}, we start at the earlier AGB phase, and use a stellar evolution code to obtain an evolved AGB model.

In order to obtain the structure of the star which is important for determining the structure constants $A$ (eq. \ref{eq:A}) and $Q$ (eq. \ref{eq:k_AM}) that are required for the tidal force calculations (eq. \ref{eq:ftf}, \ref{eq:sigma} and \ref{eq:fqd}), we use the version 10000 of the \texttt{MESA} stellar evolution code \citep{Paxtonetal2011,Paxtonetal2013,Paxtonetal2015,Paxtonetal2018} to develop a star with a zero age main sequence (ZAMS) mass of 
$M_{\rm ZAMS} =1.3 \rmModot$, to the post-AGB stage and use its properties at that time.
We obtain an evolved AGB star with mass $M_1=0.6 \rmModot$, radius of $R_1=139 \rmRodot$, luminosity of $L_1= 2230 \rmLodot$ and effective temperature of $T_{\rm eff}\simeq 3370 \K$.
The envelope mass is $M_{\rm env}\simeq0.57 \rmModot \simeq 0.95 M_1$ and its density profile is  sharply decreasing towards the center.

We use the \texttt{MESA} results in setting a model for the dynamical calculation described in section \ref{sec:evolution}.
We start the simulation when the star is still on the AGB, with the parameters listed above, but $M_1=1 \rmModot$
We take the resulting \texttt{MESA} stellar density profile and approximate it using a polytrope.
We find that it is best approximated by a polytropic density profile with polytropic index $n=4$.
The high polytropic index essentially means that the star has a steeper density profile at the core and shallower density profile at envelope, which is the case for the evolved star we discuss.
We use the approximated polytrope rather than the real density profile, as it is more simple to calculate the structural constants of the star $Q$ (and consequently $A$) for a polytrope than for a tabulated stellar model.
For a polytrope we can use the approximation \citep{Eglleton2006}
\begin{equation}
Q \approx \frac{3}{5} \left(1 - \frac{n}{5}\right)^{2.215}
e^{0.0245n-0.096n^2-0.0084n^3}.
\end{equation}

We take the stellar rotation to be $\mathbf{\Omega} / \mathbf{\Omega}_{\rm crit}=0.9 $, where $\mathbf{\Omega}_{\rm crit}$ is the critical angular velocity of the star.
The secondary star is treated as a point mass and we shall adopt ${M_2=0.35 \rmModot}$.
The initial binary separation is $P=322 ~\rm{days}$ which gives a semi-major axis of $a\simeq1.0 \AU$.
The initial orbital eccentricity is taken as $e=0.34$.

\begin{figure}
\centering
\includegraphics[trim= 0.4cm 0.2cm 0.4cm 0.0cm,clip=true,width=1.0\columnwidth]{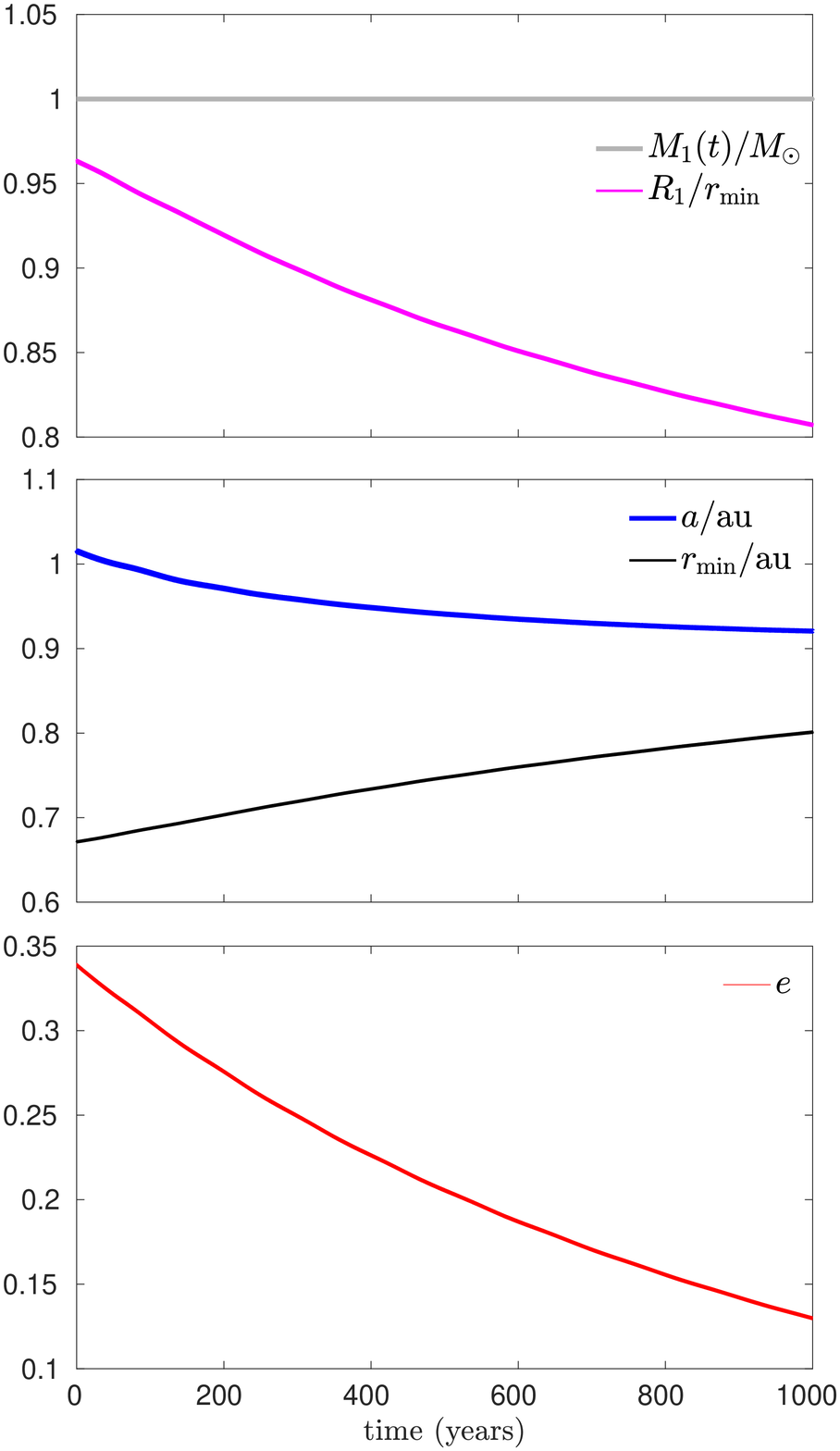}
\caption{
Results of 1\,000 years of evolution of the AGB star on its evolution path to a post-AGB star,
interacting with a secondary star through tidal force only.
The orbital evolution as a result of tidal effects alone with no mass loss and mass transfer.
\textit{Upper panel:} Mass of the AGB star and the ratio between the radius of the AGB star and the periastron distance.
\textit{Middle panel:} semi-major axis and periastron distance.
\textit{Lower panel:} orbital eccentricity. The eccentricity substantially decreases.
}
\label{fig:tidalonly}
\end{figure}

Fig. \ref{fig:tidalonly} shows the orbital evolution as a result of tidal effects alone with no mass loss and mass transfer.
It can be seen that the eccentricity decreases substantially.
It can be seen that over 1000 years the eccentricity decreased from $e=0.34$ to $e\simeq 0.13$, and after another 2000 years (out of the range shown in Fig. \ref{fig:tidalonly}) it decreased to $e\simeq0.02$.
Namely, tidal forces alone circularize the orbit in a relatively short time. 
%

\section{Periastron-enhanced mass loss}
\label{sec:Periastron-enhanced}

Next, we add mass loss and mass transfer to the calculation.
We assume that the secondary star accretes a small fraction of the mass the primary looses through an accretion disk and launches jets.
We further assume that the jets remove large amounts of gas from the envelope of the AGB star and/or from the acceleration zone of its wind \citep{Shiberetal2017, ShiberSoker2018,LopezCamaraetal2018}. Namely, we consider a grazing envelope evolution in an eccentric orbit.
We take the mass loss rate of the primary star to be ${ \dot{m}_{l1,0} = 3 \times 10^{-4} \msyr }$.
As we show below, this value, which may be typical or somewhat higher than typical post-AGB mass loss rates (as a result of the interaction with the secondary), is enough to counteract the reduction in eccentricity.

We assume that the mass loss rate of the giant star is enhanced by the jets according to a simple formula (motivated in part by the enhanced mass loss rate prescription of \citealt{ToutEggleton1988}), that depends only on the orbital separation 
\begin{equation}
\dot{m}_{l1} (r) = \dot{m}_{l1,0} \left( \frac{r}{a} \right)^6.
\label{eq:massloss}
\end{equation}
We neglect the variation in $Q$ and $A$ as a result of the mass loss of the AGB star.
Typically a few percents of the mass lost from the primary will be accreted by the secondary.
We take the accretion rate to be $5$ per cent of the mass loss rate from the giant
\begin{equation}
\dot{m}_{t} (r) = 0.05 \dot{m}_{l1} (r).
\label{eq:massaccretion}
\end{equation}
During the 1000 years that we simulate, the giant star loses $M_{\rm{ej}}\simeq 0.36 \rmModot$, bringing the giant star mass to be close to that of HD~44179 (the central star of the Red Rectangle).
During the same evolutionary time the secondary star accretes a total mass of $M_{\rm{acc}} \simeq 0.018 \rmModot$. A low mass main sequence secondary star, in the mass range of $\approx 0.3$--$0.7 \rmModot$ has a large outer convective envelope and we do not expect it to expand much while accreting a mass of $\approx 0.02 \rmModot$. If about 10 percent of the accreted mass is launched at the escape velocity of about $600 \kms$, it has has enough energy to unbind the mass that we remove.   

\begin{figure}
\centering
\includegraphics[trim= 0.4cm 0.2cm 0.4cm 0.0cm,clip=true,width=1.0\columnwidth]{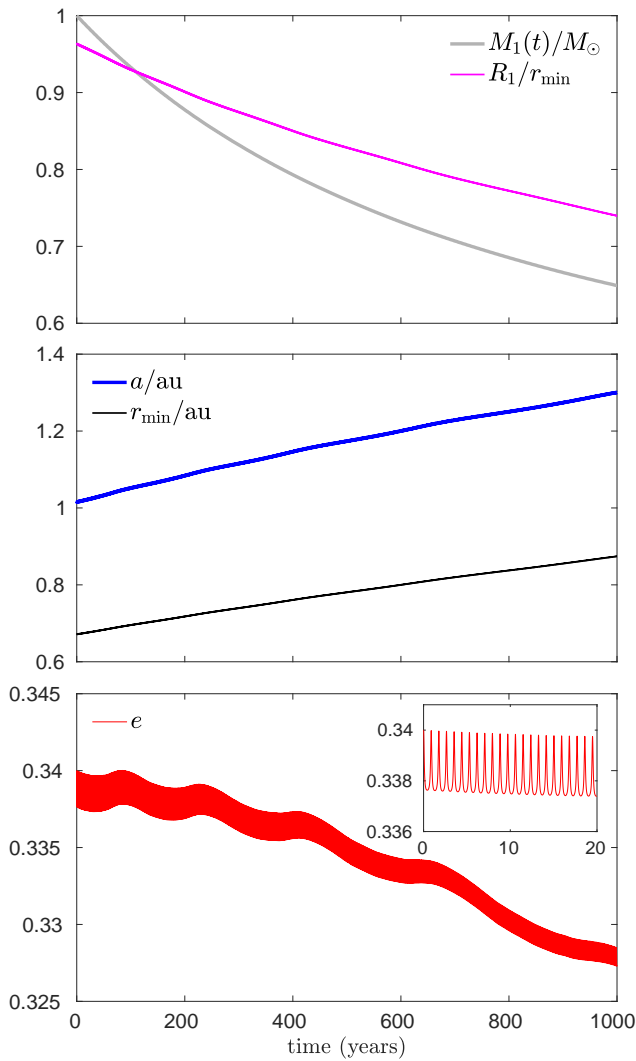}
\caption{
Same as Fig. \ref{fig:tidalonly}, but here the AGB star interacts with a secondary star through tidal forces and has an enhanced mass loss rate according to equation (\ref{eq:massloss}), 5 per cent of which is transferred to the secondary star.
We see that the value of the eccentricity stays close to its original value.
The inset on the lower panel shows the variation of the eccentricity during the first 20 years. The orbital variation makes it appear as a very thick line.
Note that the scales of vertical axes are different than those of Fig. \ref{fig:tidalonly}.
}
\label{fig:tidalandmasslossacc}
\end{figure}

Fig. \ref{fig:tidalandmasslossacc} shows the evolution of the post-AGB star and the orbital parameters over a duration of 1\,000 years.
We find that during these 1\,000 years of evolution the eccentricity did not change by much, and decreased from its initial value of $e=0.34$ to $e\simeq0.327$. 

We note that though the lines in the Fig. \ref{fig:tidalonly} and \ref{fig:tidalandmasslossacc} seem to be monotonic, there are variations, including during each orbit, which result from the varying influences of the different effects we considered here, i.e.,  tidal interaction, mass loss, and mass transfer.
The smooth line is obtained from a secular effect over many periods, each changing the orbital separation and eccentricity slightly.
We demonstrate the variation within orbits in the inset in the lower panel of Fig. \ref{fig:tidalandmasslossacc}.

The main conclusion of our parametric study is that it is possible with a reasonable set of parameters to maintain a highly eccentric orbit when we consider an enhanced mass loss rate due to strong binary interaction at periastron passages. In particular, we attribute the enhanced mass loss rate to jets launched by the secondary star, namely, to the GEE \citep{Shiberetal2017, ShiberSoker2018,LopezCamaraetal2018}.

\section{SUMMARY AND DISCUSSION}
\label{sec:summary}

We conducted numerical calculations to show that substantially enhanced mass loss at periastron passages can counteract the circularization effect of the tidal interaction. Consequently, the binary system can maintain its initial high eccentricity. Such a very high mass loss rate at periastron passages requires an extra energy source to that of the orbital energy. The extra energy can come from the accretion of mass from the giant envelope or its slow wind close to its surface onto a more compact secondary star. The secondary star energizes the outflow by launching jets from the accretion disk. 
This process of mass removal by jets as the secondary star grazes the giant envelope is termed the GEE (for grazing envelope evolution).  

To reveal the role of the GEE in maintaining the high eccentricity we compared a numerical calculation with tidal interaction only (Fig. \ref{fig:tidalonly}) to a numerical calculation that includes enhanced mass loss rate at periastron passages as well as mass transfer from the AGB star to the secondary star (Fig. \ref{fig:tidalandmasslossacc}). We found that when we included only tidal interaction the binary system underwent rapid circularization, i.e., the eccentricity rapidly decreased (lower panel of Fig. \ref{fig:tidalonly}). On the other hand, when we included the expected effects of the GEE we found that the system maintained its eccentricity. 

The high average mass loss rate of $\langle {\dot{m}_{l1} \rangle = 3.6 \times 10^{-4} \msyr}$ that we had over the 1\,000 years of our calculation is driven mainly by the jets that the secondary star launches as it accretes mass through an accretion disk. This serves as the extra energy source that is required to eject part of the envelope at periastron passages and by that to maintain the high eccentricity. 
The removal of the envelope by the jets operates via a feedback cycle that includes a positive part and a negative one. 
In the positive part of the feedback cycle the jets themselves remove mass and energy from the vicinity of the secondary star, and by that reduce the pressure there and allow a high accretion rate (e.g.,  \citealt{Shiberetal2016, Chamandyetal2018}).    
In the negative part of the feedback cycle the jets remove mass from the ambient medium, the envelope or the wind, which serve as the reservoir of accreted gas. This in turn reduces the  accretion rate and hence the jets' power (e.g., \citealt{Sokeretal2013, LopezCamaraetal2018}). This ensures that the jets do not remove too much envelope, more than exists there. 

Our proposed model can also explain other systems in which the eccentricity is larger than expected.
Using Kepler observations, \cite{Kawaharaetal2018} have recently discovered a number of binary systems with $\simeq 0.6 \rmModot$ WDs and a stellar companion with moderate eccentricities of $\simeq 0.06$--$0.2$, and typical orbital separation of $\simeq 1.4$--$2 \AU$.
According to traditional evolutionary calculations the tidal interaction in such evolved binary systems should have acted to circularize the orbits. 
One of the ideas that \cite{Kawaharaetal2018} suggest to overcome this difficulty is a scenario with insufficient time for the tidal interaction to reduce the eccentricity to very low values. The model we studied here, that is based on periastron enhanced mass loss rate in a GEE, does not require a limited interaction time, and can account for the eccentric orbits in those systems.

The scenario we studied here can undoubtedly work to maintain eccentric orbits. 
The establishment of the GEE mechanism to maintain high eccentricity requires more calculations.
First are studies of a much larger parameter space that will be based on hydrodynamical simulations of the GEE in eccentric orbits. A second type of calculations is a comprehensive binary-population synthesis to determine the likelihood of our suggested scenario as the dominant one in different kinds of AGB and post-AGB binary systems, in particular post-AGBIBs.

\vspace{0.3cm}
\section*{Acknowledgements}
We thank Kento Masuda and an anonymous referee for helpful comments.
This research was supported in part by the Israeli Science Foundation and the Pazi foundation.
AK acknowledges support from the Authority for Research \& Development in Ariel University and the Rector of Ariel University.
This work was supported by the Cy-Tera Project, which is co-funded by the European Regional Development Fund and the Republic of Cyprus through the Research Promotion Foundation.

\label{lastpage}
\end{document}